# Astronomy Identity Framework for Undergraduate Students and Researchers

Zachary Richards
*Department of Earth & Physical Sciences, York College, City University of New York,
94-20 Guy R. Brewer Blvd, Jamaica, NY 11451, USA*
*Institute for STEM Education, Stony Brook University, 092 Life Sciences, Stony Brook, NY 11794-5233, USA*

Angela M. Kelly[1]
*Department of Physics & Astronomy and Institute for STEM Education,
Stony Brook University, 092 Life Sciences, Stony Brook, NY 11794-5233, USA*

This research was a qualitative transcendental phenomenological exploration of astronomy identity formation among astronomy majors and physics majors engaged in astronomy research. Participants ($N$=10), all of whom identified with traditionally marginalized groups in astronomy, were recruited from two large universities in New York State at different stages in their undergraduate careers. *Social cognitive career theory* and the *physics identity framework* conceptually guided the analysis of astronomy identity for undergraduate majors and undergraduate astronomy researchers by exploring participants' interest in, choice to study, and persistence in astronomy. Themes related to astronomy interest were popular culture and directly observing astronomical phenomena, while astronomy choice and persistence were facilitated by experiences in introductory coursework, recognition from faculty, and socializing with peers. The emergent astronomy identity framework was characterized by six distinct yet interrelated constructs: (1) interest, typically rooted in observing naturally occurring phenomena and engaging with popular culture; (2) recognition from peers, experts, and families; (3) peer socialization; (4) competence; (5) sense of belonging; and (6) astronomy career expectations. Implications from this research provide insights on factors that influence undergraduates in four-year colleges to study astronomy, and how students' past experiences lead to a natural interest in astronomy that may be fostered in secondary and post-secondary contexts. Findings suggest departments and institutions may facilitate the accessibility of astronomy at the collegiate level by promoting a more inclusive astronomy community, fostering interactions with astronomy faculty and graduate students, providing opportunities for undergraduate research, and communicating expectancy for astronomy-related future careers.

## I. INTRODUCTION

There is a need to understand the self-concept and motivations of undergraduate students who choose to study and perform research in astronomy. In the 2021-22 academic year, there were 870 bachelor's degrees awarded in astronomy across combined and standalone astronomy departments in U.S. higher education [1]. Since the number of astronomy majors in U.S. higher education is relatively low, the majority of astronomy education research focuses on students in general education astronomy courses [2,3], which typically enroll non-science majors [4]. Research on astronomy major choice and persistence at the undergraduate level has been limited [5,6]. Although there has been significant research on the physics identity of undergraduate students [for example, 7-10], there has been minimal research on astronomy identity. This work seeks to address this void in astronomy education research with a sample of ten demographically diverse undergraduates who majored or performed research in astronomy during the 2022-23 academic year.

---
[1] angela.kelly@stonybrook.edu

There are similarities and differences between physics and astronomy disciplines and epistemological approaches. Both are disciplines that experience disparate underrepresentation of women and racialized ethnic minorities [1]. With physics, students are often able to physically interact with systems in their natural world. Conversely, students studying astronomy cannot physically interact with the systems they are studying, often relying on large-scale data sets and computers to conduct their observations and research. Given the inherent differences between physics and astronomy study, and the need for producing more graduates in the field, the development of an astronomy identity framework may allow for increased recruitment and retention in this discipline.

## II. LITERATURE REVIEW

### A. Physics and astronomy identity

Although astronomy identity has not been addressed substantively in previous literature, there has been research on factors contributing to physics identity. Hazari et al.'s [7] physics identity framework includes factors such as physics performance, competence, recognition from others, and interest. Experiential aspects include class participation, teacher encouragement, and regular discussions of current science topics, all of which have been shown to positively affect physics identity [7]. This framework is rooted in Carlone & Johnson's [11] science identity framework, which consists of performance, recognition, and competence. The physics identity framework [7] expanded upon this by adding interest as a construct.

Career expectations have also been shown to relate to physics identity. In a study exploring 15,847 responses from the *Outreach Programs and Science Career Intentions* survey, students who intrinsically valued career outcomes had increased physics identity [12]. Institutions may encourage physics and astronomy students to participate in undergraduate research, but it is often not required. Students who perform research often build a sense of physics identity [10], which may change over time through an increased sense of ownership [13].

### 1. Gender-Based Identity Studies

Research has found significant gender disparities in physics identity, with young women more likely to experience less physics self-efficacy and self-concept [14,15]. External factors such as recognition and validation have been identified as more impactful for women in high school, undergraduate, and graduate physics when compared to men [10,16]. For women, interest may develop because they received recognition from both teachers and parents [17,18], which may influence their performance and sense of competence [19]. Inclusive teaching practices and active teacher recognition have been identified as powerful strategies for building a sense of physics identity among women [20]. One such strategy is explicitly addressing gender-based physics underrepresentation in high school physics classrooms – this communicates a sense of caring and encouragement [7]. The intersectional physics identities of women of color and LGBTQ+ students have also been explored, with research finding they strengthen identity when overcoming gender and sexual orientation biases [21].

### 2. Physics Identity and Racialized Minorities

Physics identity research has also been explored for racialized minority groups. Hyater-Adams et al. [8,9] devised a critical physics identity framework for African-American physics students. It built upon Hazari et al.'s [7] physics identity framework, noting that some ethnic and gender minorities tend have more negative experiences in physics and experience higher levels of imposter syndrome [8,9]. Research on persistence of graduate students in physics and astronomy has noted imposter syndrome to be a cause of attrition in the astronomical field for both women and men [6]. Research has also found that African-American students are less likely to participate in activities which contribute to building a sense of physics identity [22], which is consequential since participation in astronomy outreach programs has the potential to increase African-American students' sense of astronomy identity [23].

### 3. Astronomy Identity in Younger Students

Although research on astronomy identity is limited, a recent study sought to develop an astronomy identity framework for middle school-aged students [24], identifying four constructs: (1) interest; (2) recognizing the utility value of astronomy; (3) confidence in performing astronomy tasks; and (4) astronomy conceptual knowledge. This framework noted that girls tend to experience an increased sense of astronomy identity that originates from interest, while overall astronomy identity tends to decrease between $5^{th}$ and $9^{th}$ grades. Interest is often developed by engaging in amateur astronomy and observing natural phenomena [24]. This may translate to older students in related fields. When considering physics identity of students in higher education, both undergraduate and graduate students who participate in public outreach or informal programs with the public may feel more like a physicist because of public recognition [25].

### B. Internal motivations for studying physics and astronomy

Students often decide they want to study physics before college because they have a passion for physics and or an emotional connection to it [26,27]. Since astronomy is often not required in high school [28], students may take it as an elective or learn some astronomy topics in their physics or Earth science courses. Research has found that students are generally interested in astronomy during elementary and middle school, yet they tend to lose interest by the time they finish high school [29,30]. Students may build interest in physics and astronomy through informal outreach activities [22,29], and this interest has been correlated to identity development [7]. Some informal activities that affect physics identity are observing stars, tinkering, and talking science with other people [22]. For example, students who observed stars in an astronomy club were 2.7 times more likely to be interested in an astronomy career by the end of high school [29]. Underrepresented groups are less likely to participate in events related to building physics identity because it is often thought that one must be gifted mathematically to study or participate in such events, which negatively impacts their participation in these subjects [31,32].

While many students decide to study physics and astronomy before entering college [26,27], some decide after matriculating in college. Introductory level courses may inspire students to pursue the physics and/or astronomy major [33]. McCormick et al. [34] interviewed five female graduate students in physics and astronomy about what influenced their academic path. The participants indicated their interest increased during their introductory physics or astronomy classes, noting the influence of skilled pedagogy [33,34].

Research and laboratory experiences in the classroom may have an impact on a student's astronomy identity, fostering interest in the subject. Research that explored *Course-Based Undergraduate Research Experiences* (CURE) in astronomy found that students improved their sense of competence to perform astronomy research, developed a sense of belonging in the astronomical community, and saw themselves as both scientists and astronomers [35]. In an introductory astronomy course where students learned how to operate telescopes and perform their own astrophotography, 71% of the students reported an increased interest in astronomy and 89% of the students in this class enjoyed the course due to the hands-on nature and practicality of the course [36].

### C. Sociocultural motivation for studying physics and astronomy

Popular culture has the potential to influence one's career choices, and there is a proliferation of astronomy-related media that are easily accessible for students of all ages. Students may first experience physics and or astronomy from popular culture depicting astronomers, physicists, and scientific discoveries [37,38], which often facilitates aspirations to major in these fields. Students who reported watching or reading science fiction were 1.5 times more likely to be interested in a career in astronomy by the end of high school [29].

A sense of belonging, familial influence, community, and identity have been identified as key factors in fostering continued interest and success in college for physics majors. A sense of belonging facilitated by a supportive physics community may improve physics recruitment and retention [38]. Sense of belonging also relates to students' self-efficacy in physics [14]. Students' career paths are often influenced by their parents and teachers [18]. Although parents often rank the mathematics and science abilities of boys higher than girls [39], girls receive more encouragement to pursue physics than boys in high school [40]. Parents may encourage students to study physics by including them in their own physics hobbies [18]. Recognition from teachers also influences students' choices. Avraamidou [17] explored the accounts of three female high school students, noting that teachers directly recognized students' talents and encouraged them to pursue physics in college. Such recognition has been shown to contribute to the development of career aspirations in science, technology, engineering, and mathematics (STEM) fields [41].

### III. CONCEPTUAL FRAMEWORK

There is a substantial gap in the literature regarding astronomy identity and why students choose to study astronomy. The present study utilizes Lent et al.'s social cognitive career theory [42] and Hazari et al.'s [7] physics identity framework to identify potential mechanisms for understanding how undergraduate students choose and persist in the study of astronomy. These conceptual approaches are grounded in social cognitive and environmental constructs that influence academic behaviors and the formation of career aspirations. In examining the astronomy-specific context through inductive analysis, a novel explanatory framework that draws upon social cognitive career theory and the physics identity framework will be developed.

Social cognitive career theory articulates factors that influence an individual's choice to pursue a particular field [42]. This theory draws upon Bandura's seminal work in social cognitive theory, which suggests that an individual's self-efficacy influences academic behavioral intentions, which are also guided by outcome expectations [43]. Sociocultural influences have reciprocal interactions with self-conceptions, motivations, and actions [44].

Lent et al. [45] built upon Bandura's work by examining social cognitive influences on science-specific career choices. They identified prior academic competence as an influence on interest and vocational intention, along with outcome expectations, or the "belief that a given behavior will lead to a particular outcome" [45 p. 424]. Lent et al. [46] had previously suggested connections among self-efficacy, interest, and motivation, which contribute to persistence in a discipline. Intrinsic and external motivations serve as mechanisms for individual choices regarding academic pathways and desired careers [47]. Intrinsic motivation is one's tendency to seek challenges to improve learning and knowledge, and extrinsic motivation is when an individual's actions are facilitated by external pressures [47]. The present study explores astronomy students' self-efficacy, interest, and motivation for participating in astronomy through research and/or an academic major, while identifying both internal and external influences.

The physics identity framework cites recognition, interest, performance, and competence as contributing factors to one's physics identity, a complex construct involving a student's self-identification with regard to the discipline [7]. Recognition is the acknowledgment by peers and/or physics experts as having physics competence. Interest is one's desire to understand physics. Competence is the belief that one is able to understand physics, and performance is the ability to complete required physics tasks. Interest, competence, and performance may be considered intrinsic motivations and recognition may be considered an external motivation. Each of these constructs is dynamically interrelated. Research has shown that students who see themselves as physicists are 3.1 times more likely to consider a career in physics [7].

Interest in a particular subject may come from self-efficacy in that particular discipline, which may be rooted in academic performance and the belief that one can accomplish a task [42]. Self-efficacy and interest may be influenced by external factors such as recognition from peers and faculty [42]. A student who performs well might receive recognition, leading to validation that may influence choice of major [7]. Previously, competence was defined by academic performance [11], however, Gonsalves [48] modified this construct by identifying a distinction between performance and competence. In a study exploring what factors allow female doctoral (astro)physics students to gain recognition, competence was redefined through specific skill sets, for example, the ability to problem solve, teach, or communicate with physics jargon rather than perform academically [48].

Engagement in research early in one's undergraduate career may be instrumental in STEM career choice. Performing research allows students to help build an identity as a scientist [11]. Specifically, participating in research early in the academic pipeline allows students to bond with other members of the laboratory through shared experiences, which may contribute to a sense of belonging. In a study examining how internal and external factors affected physics identity, undergraduate research played a pivotal role [10]. Through their research, students became part of a community and received recognition from others which strengthened their science identity [11]. Being part of a research group and experiencing recognition from faculty allowed women students to become socially integrated in their department [10].

Figure 1 represents the triadic reciprocity among self-efficacy, behavioral influences, and career expectations with regard to student participation in astronomy studies and research. Both self-efficacy and competence describe one's belief on how well they can perform specific tasks related to astronomy. Recognition relates to behavioral influences. The more positive validation a student receives, the more socially engaged they may become in their department. Socialization between students and faculty has the potential to influence behavior and identity through recognition. Career expectations may be formed through performing research and interacting with faculty and graduate students. During this important and formative time, students may also learn skills necessary to be an astronomer, furthering their career expectations. These constructs informed interviews with students as a novel astronomy identity framework was inductively generated.

FIG. 1. *Potential constructs in an astronomy identity framework* (adapted from [7,42,45-47])

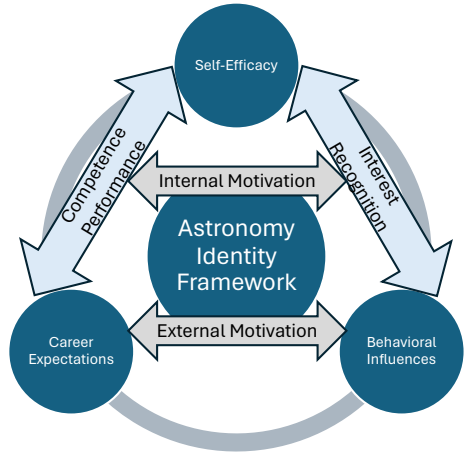

The present study identifies key factors that influence astronomy identity, examining students' choices through the lens of social cognitive career theory [42,45-47] and the physics identity framework [7] to propose a new astronomy identity framework for undergraduate students. An astronomy identity framework may shed light on how to increase the number of astronomy majors and improve the diversity and inclusion of the astronomy discipline. The present study aims to answer to answer the following research questions:

1. What factors influenced undergraduate students' decision making in studying astronomy and/or participating in astronomy research?
2. What factors influenced persistence in astronomy research and/or the astronomy major?
3. How might an astronomy identity framework be differentiated from prior physics and science identity frameworks?

## IV. METHODS

### A. Research design

A qualitative research design was employed with a transcendental phenomenological approach since it was necessary to understand multiple individuals' experiences in studying astronomy [49]. A *transcendental phenomenological* study focuses on interviewing up to 15 individuals with a shared experience and reduces data to specific quotes to provide descriptions of their experiences [49]. Common themes may be identified and explored to understand how an event was experienced by the individuals. In this case, it was how and why undergraduate students chose to study astronomy, their experience in astronomy research and majors, and how these factors contributed to their astronomy identity.

### B. Context and participants

To understand why students chose to major and/or perform research in astronomy and how their astronomy identity formed, participants were recruited from two higher-education institutions in Southeast New York State. The first was a suburban research university with an enrollment of 26,000 students, and the second was an urban university with an enrollment of 23,000 students. These institutions were chosen to provide a purposefully diverse sample of study participants. Institution 1 offered both the astronomy major ($N=54$ students in 2024) and minor ($N=5$); 37% of the astronomy majors also majored in physics. Institution 2 only had a physics major, however, many faculty were active in astrophysics research. Institution 2 had 18 students who were physics majors performing research in astronomy. Participants in this study from Institution 1 were astronomy majors, double majors in physics and astronomy, or astronomy minors. Those from Institution 2 were physics majors performing research in astronomy.

The characteristics of the study participants (identified by pseudonyms) are shown in Table 1. They were recruited through e-mails to undergraduate astronomy majors and researchers and provided voluntary informed consent (IRB approval #2022-00287). The participants ($N=10$) included six women and four men. Six participants were majoring or minoring in astronomy; four others majored in physics or Earth and space science and did research in astrophysics. Seven participants were juniors or seniors. Five participants self-identified as Hispanic, while the others were South Asian, Middle Eastern, and Caucasian. Notably, all participants identified as belonging to at least one demographic group underrepresented in astronomy.

TABLE 1. *Characteristics of participants*

| Participant | Institution | Academic Year | Major | Astronomy Research | Gender | Ethnicity |
|---|---|---|---|---|---|---|
| Alberto | 1 | 3 | Physics; Astronomy Minor | No | Male | Hispanic |
| Alfonso | 1 | 2 | Astronomy | No | Male | Hispanic |
| Leticia | 2 | 4 | Physics | Yes | Female | Hispanic |
| Diego | 2 | 4 | Physics | Yes | Male | Hispanic |
| Alejandra | 2 | 4 | Physics | Yes | Female | South Asian |
| Luis | 1 & 2 | 3 | Physics & Astronomy | Yes | Male | Hispanic |
| Sondra | 1 | 3 | Astronomy | No | Female | Caucasian |
| Olivia | 1 | 1 | Earth/Space Science | No | Female | Caucasian |
| Cathy | 1 | 4 | Astronomy | No | Female | Caucasian |
| Dahlia | 1 | 2 | Physics & Astronomy | Yes | Female | Middle Eastern |

### C. Coding process

The coding process was conducted in three phases following procedures outlined by Saldaña [50] and Creswell [49]. A priori coding with elements of grounded theory were utilized [49]. First, open coding was used to identify key phrases in the transcripts that aligned with the research questions and conceptual framework. Open codes were then refined to have a better representation of the data. In the second phase, axial coding, the researchers utilized *versus coding* to identify nuanced information regarding positive, negative, and neutral influences on participants' interest in astronomy, choice to study astronomy, and astronomy identity [49,50]. Two of the general codes did not have axial codes. Four general codes contained axial codes relating to specific times, phenomena, or sense of identity. The axial codes for ten of the general codes contained versus coding, which were chosen to compare the influences of specific events on students' astronomy interest. Specific events were analyzed and identified as having a positive, negative, or neutral influence on a student's decision to study astronomy. In the third iteration, thematic coding took place. Thematic coding allowed for the organization of codes into groups to identify recurring patterns that built an explanatory narrative that built upon yet was differentiated from existing science identity frameworks [50].

Transcripts were first coded independently by the two researchers and recoded with extensive discussion between the authors six times to reach interrater reliability of 90% [51]. This method is sufficient when there are large amounts of codes [52]. A total of 38 open codes were identified. Codes are listed in Table 2, along with the frequencies of their identification.

### D. Validity and trustworthiness

Multiple measures were taken to insure the validity and trustworthiness of the qualitative methods. Since a transcendental phenomenological approach was employed, the participants were recruited through criterion sampling [53]. Students were selected from a population of suburban or urban undergraduates who studied or performed research in astronomy, since it was presumed they had similar experiences with and knowledge of astronomy [53,54]. The site variation was intentional to recruit students from diverse backgrounds.

During interviews, member checking allowed participants to add credibility by confirming their responses [55]. The interviewers selectively repeated participants' responses so they would reiterate and confirm their meaning. Interviews were coded with an iterative process until data saturation, which occurred when no additional codes were identified [55]. The two researchers reached data saturation after coding the transcripts of seven participants.

An audit trail was established for this study, including documentation of the frequencies for each code (Table 2). The researchers used thick, rich descriptions to provide detailed accounts to boost credibility of the study [55]. To improve the trustworthiness, the first author acknowledges he was acquainted with three of the participants before they were recruited through formal interactions with astronomy research colleagues. He strove to minimize potential biases through extended discussions with the second author.

TABLE 2. *Open and axial codes and frequencies*

| | Codes | Alberto | Alejandra | Alfonso | Cathy | Diego | Dahlia | Luis | Leticia | Olivia | Sondra | Total |
|---|---|---|---|---|---|---|---|---|---|---|---|---|
| 1a | Early life inspirations - high school | 2 | 1 | 0 | 0 | 0 | 0 | 0 | 1 | 1 | 0 | 5 |
| 1b | early life inspirations - beofre high school | 2 | 1 | 2 | 0 | 0 | 1 | 2 | 0 | 2 | 2 | 12 |
| 2a | astronomyical phenomena - direclty observed | 1 | 0 | 1 | 1 | 1 | 2 | 2 | 0 | 1 | 1 | 10 |
| 2b | astronomical phenomena pop culture | 1 | 1 | 0 | 0 | 1 | 0 | 2 | 0 | 2 | 0 | 7 |
| 3a | parents/family influence - positive | 1 | 1 | 2 | 1 | 1 | 1 | 0 | 2 | 2 | 1 | 12 |
| 3b | parents/family influence - negative | 0 | 0 | 0 | 0 | 1 | 0 | 0 | 1 | 0 | 0 | 2 |
| 3c | parents/family influence - neutral | 0 | 0 | 0 | 0 | 0 | 0 | 0 | 1 | 0 | 0 | 1 |
| 5a | mentor influence - positive | 0 | 0 | 0 | 0 | 0 | 2 | 4 | 2 | 2 | 0 | 10 |
| 5b | mentor influence- negative | 0 | 0 | 0 | 0 | 0 | 0 | 0 | 1 | 0 | 0 | 1 |
| 6a | Recognition from teacher/mentor - positive | 0 | 0 | 0 | 0 | 1 | 0 | 1 | 0 | 0 | 0 | 2 |
| 6b | ognition from teacher/mentor - megative | 0 | 0 | 0 | 0 | 1 | 0 | 0 | 1 | 0 | 0 | 2 |
| 7a | Astronomy Community - positive | 0 | 0 | 1 | 0 | 2 | 0 | 5 | 2 | 0 | 0 | 10 |
| 7b | Astrnomy - Community - negative | 0 | 0 | 0 | 0 | 0 | 0 | 0 | 1 | 0 | 0 | 1 |
| 8a | research - positive | 0 | 1 | 0 | 0 | 3 | 1 | 7 | 5 | 2 | 0 | 19 |
| 8b | research - negative | 0 | 0 | 0 | 0 | 2 | 1 | 0 | 0 | 0 | 0 | 3 |
| 8c | research - neutral | 1 | 0 | 0 | 0 | 1 | 0 | 0 | 0 | 1 | 1 | 4 |
| 9a | socialization - positive | 0 | 0 | 1 | 3 | 3 | 5 | 5 | 2 | 5 | 1 | 25 |
| 9b | socialization - negative | 0 | 0 | 1 | 2 | 0 | 0 | 0 | 0 | 1 | 0 | 4 |
| 9c | socialization - neutral | 0 | 0 | 1 | 0 | 0 | 0 | 1 | 0 | 0 | 0 | 2 |
| 10a | gender/ethnicity - positive | 0 | 1 | 0 | 0 | 0 | 0 | 0 | 2 | 0 | 0 | 3 |
| 10b | gender/ethnicity - negative | 0 | 1 | 0 | 0 | 3 | 1 | 0 | 2 | 2 | 0 | 9 |
| 10c | gender/ethnicity - neutral | 0 | 1 | 0 | 2 | 1 | 0 | 2 | 0 | 0 | 0 | 6 |
| 11a | performance - postive | 0 | 1 | 0 | 2 | 2 | 2 | 1 | 1 | 1 | 3 | 13 |
| 11b | performance - negative | 0 | 0 | 0 | 0 | 0 | 0 | 0 | 0 | 0 | 6 | 6 |
| 11c | performance - neutral | 0 | 0 | 0 | 0 | 0 | 1 | 0 | 0 | 0 | 1 | 2 |
| 12a | physics lecutre/lab influence - positive | 2 | 3 | 0 | 1 | 5 | 1 | 1 | 1 | 0 | 0 | 14 |
| 12b | physics lecutre/lab influence - negative | 0 | 0 | 0 | 1 | 0 | 0 | 0 | 1 | 0 | 0 | 2 |
| 12c | physics lecutre/lab influence neutral | 2 | 3 | 0 | 0 | 1 | 0 | 1 | 0 | 0 | 0 | 7 |
| 12d | physics lecutre/lab influence - introducotry | 1 | 2 | 0 | 0 | 3 | 0 | 3 | 2 | 0 | 0 | 11 |
| 12e | physics lecutre/lab influnce - advanced | 2 | 0 | 0 | 1 | 0 | 0 | 0 | 0 | 2 | 0 | 5 |
| 13 | Interst in astronomical phenomena | 0 | 4 | 0 | 6 | 3 | 1 | 2 | 3 | 1 | 2 | 22 |
| 14a | self-efficacy/confidence - positive | 0 | 0 | 0 | 0 | 1 | 0 | 3 | 3 | 0 | 1 | 8 |
| 14b | self-efficacy/confidence - negative | 0 | 0 | 0 | 0 | 1 | 1 | 0 | 0 | 0 | 1 | 3 |
| 15a | astronomy identity - feels like an astronomer | 0 | 2 | 0 | 1 | 2 | 0 | 2 | 2 | 1 | 1 | 11 |
| 15b | astronomy identity - does not feel like an astronomer | 0 | 0 | 0 | 0 | 1 | 1 | 0 | 0 | 0 | 2 | 4 |
| 15c | astronomy idenity - competence | 0 | 1 | 0 | 1 | 1 | 0 | 2 | 1 | 0 | 0 | 6 |
| 15d | astronomy idenity - research | 0 | 1 | 0 | 0 | 1 | 0 | 2 | 0 | 0 | 0 | 4 |
| 15e | astronomy idenity - community | 0 | 0 | 0 | 0 | 4 | 0 | 5 | 1 | 0 | 0 | 10 |

Total Codes 278

## V. Findings

The findings are organized by the major themes identified during the coding process. First, external factors influenced interest in astronomy from family members, observing astronomical phenomena, and popular culture. Secondly, factors influencing persistence in astronomy included socialization, recognition, and relevant coursework. Third, factors related to developing astronomy identity included communities of practice, sense of belonging, and competence.

### D. Factors influencing early astronomy interest

#### 1. Observing the night sky

Students reflected on how they developed interest in astronomy early in their lives. Many discussed how their fascination with the sky facilitated their overall natural wonder and interest in science. Luis, a third-year physics and astronomy major who emigrated from Honduras as a child, described how he observed the sky without light pollution during blackouts:

*Well, I always like to talk about one of my earliest memories when I was a little kid, back in my country, the city I lived in used to have regular citywide blackouts where there would be no electricity. And so, whenever that would happen at night, me and my family would kind of just hang out outside the front porch. And if it was a really clear night, then the sky would be filled with stars, because there would be no light pollution, since the whole city had no electricity. So, just that view of the stars, I guess, would be one of the first things that kind of got me interested.*

For Luis, the awe and wonder of the visibility of stars while in darkness was an experience he could share with his family that permeated his early memories. Notably, an energy cost-saving measure led to advantageous conditions for these observations. Cathy, a fourth-year astronomy major, recalled similar experiences with her father, who would frequently use a telescope to teach his daughter about celestial bodies in the night sky:

*I do like observing with my dad because he had a telescope. He got another one and a newer model at the time. He could, like, look at the moon and stuff like that. It was cool. That kind of benefited because you're like, "Oh" – I liked doing that. I guess to me, like, astronomy is… almost like the final frontier in science and stuff like that because there's still so much to know and what research to be done.*

Both Cathy and Luis were drawn to making astronomical observations at an early age in the company of their families. The sky presents a natural view of celestial phenomena that they found fascinating, inspiring, and relevant to their everyday lives. In terms of astronomy identity, lifelong, universal access to visible phenomena may allow students to feel a connection with the discipline before formal study.

#### 2. Popular culture

Students also reported popular culture influenced their interest in astronomy. Research has indicated that popular culture has the potential to act as a first learning experience for individuals, which may influence their choice to study astronomy [37]. Diego, a first-generation college student pursuing his second bachelor's degree in physics, was participating in research on exoplanets and cosmology. He recalled times in his childhood when he watched popular culture icons on television, which piqued his interest in astronomy:

*Carl Sagan is probably my favorite human being to have ever existed. So, I grew up just*

*seeing him sometimes on TV, which is weird because I'm not even that old, but I would still see some of the stuff. And I thought that his quotes and the way that he represented science was really interesting.*

The way Carl Sagan spoke about science had a notable influence on Diego; his observations of relatable popular culture icons inspired fascination and a desire to learn more about astronomy. Luis also mentioned Carl Sagan as an influence. In his case, Carl Sagan was a role model who influenced his career aspirations and desire to pursue this discipline. Even though he knew little about the field, the presence of such a compelling figure in popular culture fostered his early sense of astronomy identity:

*Astronomy was always just something very interesting, and I want to say I think it was maybe early high school when I watched 'Cosmos' by Carl Sagan, that's what made me realize, like, 'Yeah, I want to be an astronomer.' … I didn't know exactly what that entailed back then, but I just knew that I wanted to do something with astronomy.*

For both Diego and Luis, popular culture was a key early influence on their astronomy interest and career aspirations. Televised programs presented astronomy as a relevant, authentic scientific pursuit, while introducing role models who sparked wonder in individuals by making astronomy simple and accessible.

### B. Factors influencing choice and persistence in the astronomy major

#### 1. Socialization

Students often mentioned how socialization with peers helped them persist through their astronomy studies. Several women astronomy students participated in a peer mentoring program for women in science and engineering. They meet weekly with a small group of 5-6 women in their major, along with a third- or fourth-year lead mentor, also in their major. Dahlia was a second-year physics and astronomy major performing research. When asked about how her peer mentoring group influenced her decision to study physics and astronomy, she commented on the shared struggles with academic/life balance and performing well in coursework:

*I cannot stress how much it helped me. Like, it was like a family… And it really, like then slowly by slowly we got closer and closer with one another and it was great… like, when you would struggle with the lab and to know, like, your friend – she struggles with – it really makes you see it's not just [me] – like one of the girls, she's in my modern [physics] class this year and we are so, so close. Like so close because of it, because of the mentoring. And, like, we have each other 's back… I think a lot more girls kind of try to have a dual life where they tried to live their life and then they also try to study in the physics major and the astronomy major. […] And it felt so good like, it felt so good to, like, have time. Have time to do things except study… and you have a friend to relate to.*

The mentoring group allowed Dahlia to meet other students in her major and become part of an astronomy community of like-minded women. Research has suggested that physics majors often require a supportive community [38]. In Dahlia's case, she was able to share tensions, making her feel as though her struggles were typical and could be navigated with peer and mentor support. As Dahlia started to relate to more students, they saw each other as astronomers, contributing to their emerging astronomy identity and influencing their persistence in the major. This is consistent with prior research suggesting a sense of physics identity may be positively tied to persistence within the major [7,14].

Before his undergraduate career, Luis had never formally used a telescope nor was a part of an astronomy community. When asked about how the astronomy club in his department influenced his decision to study astronomy, Luis described how social relatedness contributed to his fascination with astronomy:

*It made me feel more connected with other people that were interested in astronomy, within [university]. And we didn't get to do much, but at least we got to, you know, set up some telescopes and look out into a different planet, and I think that was my first time where I finally got to see Saturn with my own eyes for the first time. And we like to say, here in [university], that Saturn is the gateway drug to astronomy.*

Participating in the astronomy club allowed Luis to feel part of a community, because he met people who shared his interest and enthusiasm for the discipline. Luis also noted observing Saturn contributed to his interest and persistence in astronomy. Both Dahlia and Luis shared that socializing with their peers made them feel like they part of a community of practice. Having these shared experiences within a major is an important factor for persistence [38].

### 2. Recognition from faculty and mentors

Participants noted receiving recognition from faculty and/or research mentors affected their choice to study and persist in their major. Luis was a junior in his astronomy major at the time of his interview. He transferred from Institution 2 to start his junior year at Institution 1. When asked about his confidence performing astronomy related work, he responded "You know, classes – it's a struggle. But when it comes to research, I feel a bit more confident." Despite initially performing poorly in his coursework, participating in research and receiving positive feedback from mentors allowed Luis to increase his confidence and persist through the astronomy major. He performed research with a mentor in galaxy structure and formation for the majority of his undergraduate studies, and maintained a collaborative relationship with his research group from Institution 2. His research group included faculty and students from multiple institutions, and he was selected for a leadership role in the group. When asked about his responsibilities as a mentor to younger students, he stated:

*Yeah, I'm one of the mentors, basically… They [faculty] always made it seem like the project that we're working on is my project, and so, any contributions, they acknowledge it… So, I'm mostly doing the main bulk of the work.*

Luis consistently received positive feedback from his mentor for the research he was conducting, which contributed to his sense of autonomy, agency, and self-efficacy. His contributions to the research project were acknowledged by his mentor, allowing others to see him as an astronomer who could perform his own independent research and mentor junior students. Independent research has been linked to building physics identity [10,13]. His leadership role as a mentor to more junior students in the research group and recognition from his mentor facilitated his determination and perseverance.

### 3. Experiences in astronomy coursework

The quality of instruction in formal classroom settings contributed to students' enthusiasm for astronomy. Cathy was a senior astronomy major who was involved in a peer mentoring group for women in physics and astronomy. Although Cathy did not participate in research at the time of the interview, she indicated she would start research in the subsequent semester. She was asked about how her astronomy courses influenced her decision to major in astronomy.

> *Honestly, the most influential right now is the class that I'm currently taking. Because I do think it is the best, like, professor I've had in a long time. But the other ones – [Astronomy Course] I think it was – was very nice. I like that course a lot, which did help.*

For her, the professor's open demeanor and pedagogical approach made the course enjoyable, which improved her interest and engagement with the discipline itself. Conversely, some students who did not enroll in a significant amount of astronomy coursework had less of an astronomy identity. For example, Alejandra was a fourth-year physics student who researched brown dwarfs. She enrolled in mostly physics coursework as opposed to formal astronomy courses. She applied to her institution because she was interested in an honors program, although they did not have an astronomy major. Alejandra was asked whether she felt capable of doing astronomy related tasks. She responded, "Let's say 6 out of 10," and further elaborated on how she thought she could improve her knowledge:

> *If I were to take more courses on maybe physics, and math, and astro in general. Because I haven't taken the Astro 101 course. I've only ever taken Cosmology, which was a special topics course and then a seminar course… I don't know much of the basics with orbits and what-not in Astro 101 or whatever.*

For Alejandra, the lack of relevant coursework affected her confidence in performing astronomy related tasks. Coursework can have a positive and negative affect on persistence. Skilled instruction has the potential to inspire students to choose to study physics and or astronomy, especially in introductory courses [33]. This was especially true for Diego, who was interested in astronomy since his childhood but did not know how he could become an astronomer. His introductory astronomy course was instrumental in his choice to pursue astronomy study through research, even though he perceived that one had to be brilliant to do so:

> *I've always loved space and star stuff and I – one, I had never thought that that could be an actual career that people could do… when I found out it was a career, which was at [school], because I took an astronomy class there. I thought you just had to be kind of born a genius.*

Working with faculty through coursework can have a significant impact on a student's choice to study astronomy – this often occurs through recognition and improved confidence. Research has indicated there is a common perception that mathematically intensive fields require superior cognitive skills [32]. Introductory astronomy requires basic algebra and sometimes calculus [56], and may give the impression one must excel at advanced mathematics in order to become an astronomer [57]. Such impressions may impede students from underrepresented groups from studying astronomy [57]. Many participants noted the importance of introductory astronomy coursework in deciding to study astronomy and persisting in the field. Other participants noted recognition from research mentors was a key factor relating to persistence. These findings indicate that coursework, faculty interaction, and research participation influence astronomy identity formation.

## C. Factors influencing astronomy identity

### 1. Participation in a community of practice

Students who participated in research in an active, welcoming community noted it positively affected their sense of astronomy identity. When Diego first started astronomy research, he was excited about doing work on exoplanets, although he did not feel that he belonged in the field. He noted he experienced a degree of imposter syndrome, a phenomenon when individuals feel they do not belong in a group or discipline [6,58]. He started to develop a sense of astronomy identity by performing research. When Diego was asked if he saw himself as an astronomer and the process behind developing his sense of identity he responded:

> *Not only do I see myself as an astronomer, I think it's kind of become a big part of my personal identity… I guess for the first couple of months, having worked with [mentor], I just – I thought it was way too incredible to be working on something as exciting as pointing a telescope at the sky and collecting data on stars. So, I definitely felt what I think most people feel, which is a lot of imposter syndrome… I felt like any moment the other shoe's going to drop and people are going to be like, 'No, you don't belong here.' I feel very welcomed in the community. And I feel like that is what has allowed me to see myself as more of an astronomer. So, yeah. Yeah, it's really – amongst my family, amongst my non-STEM friends, I am the astro guy. And I love it.*

Imposter syndrome is a key cause for astronomy attrition [6]. This coupled with the notion that one must be a genius to be an astronomer [32] could have stifled Diego's astronomy aspirations. He had two separate research mentors who contributed to reducing his imposter syndrome, aligning with results of Ivie et al. [6], who reported graduate school students were 1.77 times more likely to persist in astronomy if they had a secondary mentor. In addition, members of his team viewed him as an important part of their group and started to view him as an astronomer, strengthening his astronomy identity. This is consistent with research suggesting one's physics identity is influenced by how people outside the field view them [59].

Leticia was a fourth-year physics majors researching galaxy evolution. She was a first-generation college student born and raised in New York City. Half of her family immigrated from Puerto Rico. At the beginning of her undergraduate career, her family questioned her choice to pursue astronomy. Without support from her family, Leticia was at a crossroads between astronomy and physics. When she was asked about what her made her choose astronomy she answered:

> *I'll be honest. I didn't really want to do astronomy. I wasn't really sure. I just knew I liked physics… But I wasn't sure if I wanted to do the very, very small or the very, very big. So, it was either quantum or astro because I thought those were really, really cool subjects. And it actually wasn't until I got accepted into [school] that I knew I wanted to do research in astronomy… The community of [school] astro students really made me feel at home because I didn't know anybody who wanted to do physics other than me before I got into [school].*

Belonging to a community of practice attracted Leticia to astronomy. She met others with similar interests through her astronomy research, and they made her feel supported. Being involved in a peer community of practice helped develop Leticia's sense of identity as an astronomy researcher. Luis also participated in a community of practice. In his case, he was exposed to research and he started to see what it was like to be a professional astronomer. When Luis was asked whether performing research affected his desire to pursue astronomy, he responded:

*In a way, yeah, because I finally realized what is it that it takes to be an astronomer, which I didn't know back then, but I do now, that it just involves a lot of coding, a lot of computer stuff. So, yeah, it gave me a more clear view of how it is to be an astronomer… and just seeing how everything works. It definitely gave me kind of like an ease of mind, because I kind of knew what I would be – it made me see, like, what I would be doing if I were to stay in astronomy.*

Performing research gave Luis a realistic look at the day-to-day workings of a professional astronomer. These career expectations influenced his academic journey and solidified his vocational aspiration. Luis chose to continue studying to become an astronomer even though he was struggling with his courses. This resilience was influenced by gaining insight into being a professional astronomer through research, which positively affected his self-efficacy. Luis's actions are consistent with Bandura's research [43] suggesting academic behavior may be influenced by self-efficacy and career expectations.

These three participants (Diego, Letiticia, Luis) engaged in research within a community of practice. The participants discussed how their research experiences contributed to their sense of identity through collaboration and establishing career expectations. Members of their community saw them as astronomers, building astronomy identity through this recognition and mutual respect.

## 2. Sense of belonging

Participants noted a sense of belonging contributed to their emerging astronomy identity. Diego spent a significant amount of time searching for an academic place where he belonged. He was welcomed enthusiastically in his first research group and then again in his second research group. When Diego was asked about his sense of belonging in the astronomy and physics community, he responded:

*Yeah, I mean, I wrote about this in my grad school application, which I'm doing now, but I actually have never in my entire life felt like I fit in in the way that I do now when I originally switched to physics. It's like my people.*

Diego was invested in a community in which he was surrounded by others who had similar interests. Members of his community were able to discuss future steps for astronomy academic pathways, giving Diego the motivation to continue astronomical research while understanding career expectations. Dahlia also shared her feelings of belonging in the physics and astronomy major focusing on undergraduate and graduate student communities.

*I think it's more so with the grad students that I feel more connected with. Because they, like, kind of went through everything… But I think the community is always, like, always trying to be fun. Like have fun but also learn and try your best. It's always motivation to push, push, push harder and be better and always – and I see that in [Astronomy Club]. Every day you see kids studying and it makes you study.*

Dahlia felt a stronger a sense of belonging in her department because she viewed the graduate students as role models, individuals from whom she could seek assistance and advice. Dahlia's undergraduate peers built an environment where students could support each other informally, experience mutually beneficial astronomy problem solving, and maintain accountability.

A sense of belonging can be developed through communal activities and having role models. Both Dahlia and Diego were involved with communities where they had peers whom they could ask for help as well as more senior students. As their skills developed, they were able to help more junior students. Being a part of an active community often allows students to be seen as astronomers by their peers or role models. This recognition directly contributes to one's sense of belonging [14].

### 3. Competence

Multiple participants indicated confidence in their abilities affected their sense of astronomy identity. Leticia built competence through hands-on research experience. While she was confident in her abilities to perform astronomy related tasks, she did not always feel this way when faced with the failures that often occur in research. Leticia described the process of how she developed a higher level of competence through research in the face of both gender and ethnic biases:

*Failing. Figuring out why I failed and learning from it. I didn't know how to program. Learning how to program has made me way more confident in my research abilities. And just the practice of being in these academic spaces has made me more confident where it's less intimidating to be in a room full of White guys. And yeah, I think all those things, that it's – the skills and the practical uses and the practice and the time and effort I put, but also just the exposure to what academia is like so that it's less intimidating.*

Practicing astronomy-related skills such as programing and learning from her mistakes helped Leticia become more comfortable in academic spaces where latent gender and ethnic biases existed. Diego mirrored similar experiences regarding his ethnicity, stating "It felt like in a way I was representing – I'm not saying that I was, but in my head sometimes it felt like, 'Oh, I'm representing this larger [Hispanic] group.'" He has also mentioned his confidence in his abilities grew over time:

*I compare myself to myself five years ago when I never thought I could do what I'm doing now. And I also try to just keep in the back of my head recent accomplishments that I've had that kind of keep me peppy and kind of, yeah, feeling confident enough to be able to continue going.*

Diego was able to build confidence in his abilities through research, performing astronomy-related tasks, and iterative failures that became successes. Confidence allowed him to overcome his imposter syndrome and persist in astronomy. Alejandra echoed similar feelings regarding competence. While she previously stated not taking any astronomy specific courses negatively affected her sense of identity, performing research had a positive effect on her competence and identity, commenting, "Yes, I feel a little bit more capable than freshman year when I didn't know any coding and how to think like a scientist." Alejandra was able to build a sense of competence from performing research and building her capacity to complete astronomy-related tasks.

All three of these students (Leticia, Diego, Alejandra) performed research during most of their undergraduate years. This developed their skills through hands-on experiences, increasing confidence, self-efficacy, and astronomy self-concept. This indicates a relationship between competence and astronomy identity.

### VI. DISCUSSION

The qualitative data analysis in this transcendental phenomenological study contributed to an undergraduate astronomy identity framework for those who major in and do research in the discipline. Previous studies on astronomy identity formation examined how participation in course-based research affects astronomy self-concept [35], and how $5^{th}$-$9^{th}$ graders develop astronomy identity [24], however, an astronomy identity framework for undergraduates has not yet been established. This work contributes addresses the gap in astronomy education research. The work is also distinguished by the large number of women and Hispanic participants who are traditionally underrepresented in astronomy.

The present study builds upon prior physics identity research [7-10,13,20,22]. Related studies found that performing undergraduate research facilitates a sense of identify [11,13], as well as participating in community of practice events like teaching peers [7]. This astronomy identity research is differentiated from prior identity frameworks by identifying six factors within two interpretive lenses– the *internal factors* are (1) interest, typically rooted in observing naturally occurring astronomical phenomena and popular culture, (2) competence, and (3) sense of belonging; while the *external factors* are (4) socialization, (5) recognition from peers, experts, and families, and (6) astronomy career expectations. These thematic elements are described in detail below.

### A. Early interest in astronomy

The first research question examined factors that influence undergraduate students' decision making in studying astronomy and/or participating in astronomy research. Students often developed ***interest*** in astronomy at an early age through observing the night sky directly. This is often associated with positive shared experiences with their families, which suggests a cultural synergy with the scientific world. This may trigger emotional connection with astronomy sparking an interest at an early age [26,27], which is an intrinsic motivation for continuing with formal astronomy study [47]. Astronomy is accessible because no laboratory equipment is often needed; all one needs to do is look up and observe. Some students were also influenced to study astronomy through popular culture, which may act as a first astronomy learning experience for students [37, 38]. Astronomy may be addressed in popular culture and later discussed in in social situations or in the classroom acting as extrinsic motivator [47], influencing an individual to study and or participate in astronomy study and research. The ubiquitous mention of popular culture is a differentiating feature of astronomy identity.

### B. Persistence in the astronomy major

The second research question explored factors influencing persistence in astronomy research and/or the astronomy major. Students shared that ***socialization*** within the major and working with faculty were common sources of encouragement for students to persist in astronomy. Socialization through peer mentoring groups may create positive external influences on students' behavioral choices [42,47,60,61]. Socialization included but was not limited to laboratory partnerships, peer mentoring groups, and astronomy clubs. Students developed bonds and shared the stresses of their major and research tasks. During these times, students often discussed what courses to take, research opportunities, and other job or internship opportunities without faculty present. The implicit support and camaraderie often influenced the decision making of students to pursue astronomy as a major or research focus. Students sometimes joined a particular research group because of peer influence. These social experiences have been shown to be influential in shaping students' astronomy identities and sense of belonging [10,14].

Working with faculty was also influential in astronomy identity formation and persistence in the field. ***Recognition*** from faculty and mentors was an extrinsic motivation that directly affected student's behavior and one's self-efficacy, consistent with prior research [7,42,47]. Positive recognition contributed to self-efficacy; this confidence in astronomy tasks facilitates a ***sense of belonging*** in the field [7,10]. Recognition has also shown to help build physics identity, specifically for women [7]. This external validation may help students feel part of the greater astronomy community. While recognition is more influential for women [7,41], men in this study more often mentioned receiving positive recognition, indicating potential gender biases may still exist. Students also shared that peers and family members recognizing them as astronomy experts contributed to their social and academic identities.

Lastly, many students commented that relevant coursework affected their persistence in the major. Coursework often lays the foundation for astronomy, and it may positively or negatively affect students' persistence. Introductory astronomy courses may be a key recruitment tool and often cited as inspirations for studying astronomy [33,34,37]. Interactive pedagogy has been shown to increase interest in astronomy [36]. Coursework may act as an intrinsic influence since students aim to challenge themselves in their coursework to improve their own ***competence*** and desire to learn about astronomy [47]. Alternatively, coursework may act as an extrinsic influence. Students may socialize with and receive recognition from faculty and other students through asking and answering questions, participating in class discussions, and helping their peers. Recognition from faculty may be an extrinsic influence, affecting students' interest and ***career expectations,*** contributing to astronomy persistence [8,19,47].

### C. Differentiating between an astronomy and physics identity framework

The third research question addressed the operationalization of an astronomy identity framework that builds upon and is differentiated from prior physics and science identity frameworks. The physics identity framework [7] includes recognition, interest, competence, and performance. However, there are many physics and astronomy departments in addition to standalone astronomy departments [1], where the physics identity framework may inadequately describe the development of identity in undergraduate astronomy students and researchers.

The proposed astronomy identity framework consists of a dynamic interplay among interest, recognition, socialization, competence, sense of belonging, and astronomy career expectations (Figure 2). The six constructs can be categorized into internal and external influences affecting astronomy identity development. Internal influences include interest, sense of belonging, and competence; while external influences include recognition, socialization, and astronomy career expectations. Interest belongs in both physics and astronomy identity frameworks, however, the source of interest in the frameworks may be different. In the astronomy identity framework, interest often stems from early observations of the night sky and popular culture. Colantonio et al. [24] showed there was a direct link between interest and astronomy identity in children, and the elementary and middle school years may be a pivotal time frame. The vast amount of popular representations and contextualization of astronomy in films, television, social media, and books contributes to student interest in the field.

FIG. 2. *Emergent astronomy identity framework*

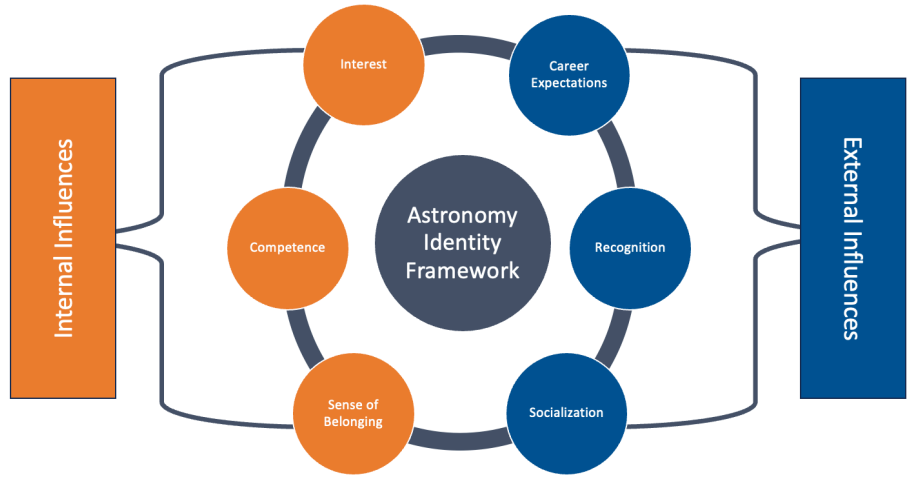

Recognition originates from prior work on science identity [11], defined as receiving validation by others as a science person. Recognition in the astronomy identity framework is being formally or informally identified by others as an astronomer [11]. This may come from faculty and mentors publicly supporting an individual as a skilled astronomy student and researcher [7]. Recognition from faculty provides validation and ability confirmation [10], contributing to one's sense of agency, confidence, and identity. Additionally, formal recognition may lead to building accurate career expectations. Faculty may see the student as an astronomer, and give them specific astronomy-related tasks which would be performed by professional astronomer.

Informal sources of recognition come from individuals outside the field, for example, family, friends, or the general public [11]. Rethman et al. [59] noted students who participated in public outreach events developed a sense of physics identity through recognition because the general public viewed them as physicists – this may translate to astronomy. It was observed that several participants started to develop a sense of astronomy identity because their family and friends saw them as astronomers.

The astronomy identity framework adopts the definition of competence as one's belief in their ability to perform astronomy related tasks [48], which is similar to other science frameworks [7]. However, the source of competence in astronomy may be more likely to originate from research experiences since astronomy major programs are far less common. Developing a sense of competence is related to receiving recognition from mentors and peers [42]. For many astronomy students, this may only be possible in an undergraduate research program.

Sense of belonging is feeling that an individual is a part of the astronomy community, which is often facilitated by participation in research [11,35]. Astronomy undergraduate researchers typically socialize with other students, faculty, and research staff. Such interactions may make one feel part of a community and socially integrate them in their departments [10]. Literature suggests students may feel a greater sense of belonging if they have higher level of competence, which may be facilitated by performing research [14,19]. If a student believes they can do what is required of an astronomer, they may feel as though they belong in the department partially from recognition from faculty and through mentoring other students. A sense of belonging is also built through socializing with peers in their community, allowing students to discuss classes, research problems, and tensions.

Socialization is a new construct in the astronomy identity framework and is intertwined with a sense of belonging. Socialization with peers and faculty occurs in formal and informal settings such as performing research [10], participating in astronomy clubs [29], joining peer mentoring groups, or participating in communities of practice. Eight of the ten participants cited socialization as an important influence in studying astronomy. Socializing acts as an external behavioral influence as students may make decisions based on what faculty and peers do or say [42,47]. Socializing with faculty may give students a clear representation of the day-to-day life of a professional astronomer. Socialization is also tied to receiving recognition. In communities of practice, students may help one another. Through these reciprocal processes, students may view themselves as astronomers in practice.

The last construct related to building astronomy identity is career expectations, which are informed by the notion that a particular behavior will lead to a desired outcome [42]. Career expectations are how students perceive particular vocations in terms of day-to-day tasks, personal values, and potential for advancement. Students' career expectations can be developed by performing astronomy research and socializing with faculty. The behaviors of faculty and peers have the potential to act as an external influence [42,47]. Performing research sets the expectations of what being a professional astronomer is like by developing skills, normative assessments of competence, awareness of career opportunities. It has been shown performing research positively affects students' astronomy identity and helps develop the skills required to be an astronomer [35]. As students hone their skills through research, they become more competent and may see themselves as astronomers. Career expectations may be considered the most influential construct in an astronomy identity framework. If a student has started to build a sense of astronomy identity through the other five constructs, but does not see themselves doing astronomy-related tasks in their careers, they may not develop a sense of astronomy self-concept.

### D. Implications

There are several implications from the qualitative analysis presented in this research. Astronomy interest may be facilitated at early ages through public outreach programs since most students will not take a formal astronomy class in high school [22,28]. Informal public outreach events have the potential to spark interest and create an emotional response with parents to influence students to study astronomy in college. Informal outreach programs should be expanded and encouraged to inspire the next generation of astronomers [23]. This may be through university partnerships or informal science institutions. Professional learning programs for precollege science educators may encourage them to incorporate astronomy education resources in their science instruction. Informal science institutions have been effective in providing interactive exhibits for students to learn science that is difficult to access in traditional classrooms [62,63].

Introductory astronomy courses are important for recruiting students in physics and astronomy departments [33,34]. It is advised that interactive pedagogies should be used in the introductory courses to foster interest and recruit students to the astronomy major [33]. Some hands-on approaches for teaching astronomy classes include the use of telescopes and course-based research projects. These activities have been shown to increase interest in astronomy and students' astronomy identity [35,36]. Astronomy programs may make research a required component of the major, leveraging the positive outcomes identified by prior studies [10,13]. By performing research, students often receive recognition from peers and faculty, increase their competence, and develop a sense of belonging through socialization. This experience often results in a positive impact on students' astronomy identity, which may inspire them to persist in the major. Socializing with astronomy faculty is important and should be done more frequently to give students a better idea of what they can expect for a career in astronomy, which may set realistic and obtainable goals for students' futures.

### E. Limitations

This study has several limitations. First, there was a relatively low number of participants as there were not many astronomy majors and students performing astronomy research. Not every participant was an astronomy major, and each participant interacted with the field of astronomy in different ways. Second, participants for this study were selected from higher education institutions in New York State – it is possible that some of the research may not translate to other geographic regions. Third, interview transcripts were the only source of data. Including other sources such as observations and course artifacts has the potential to increase the validity and reliability of our analysis. The retrospective interviews discussed events which occurred in the participant's past and therefore could not be observed. Fourth, the researchers are physics education and astronomy educators at higher education institution. The coding of interview transcripts may have been affected by the researchers' biases.

### F. Conclusions

Research on astronomy identity is limited, and the research presented in this paper developed an astronomy identity framework by examining mechanisms related to interest and persistence in astronomy. The development of an identity framework may provide insight on fostering interest and persistence in astronomy. Interest in astronomy forms at an early age, usually from observing the night sky or through popular culture. Persistence in an astronomy major often consists of working with faculty and socialization within the major. Astronomy identity consists of interest, recognition, competence, socialization, sense of belonging, and career expectations. Each construct helps develop students' sense of astronomy identity. Astronomy education researchers might utilize this work to expand and diversify participation in the field.

### ACKNOWLEDGMENTS

This work was supported by the National Science Foundation [I-USE #2142587].

## APPENDIX: INTERVIEW PROTOCOL

1. Demographics
   - What college(s) have you attended?
   - Why did you choose your college(s)?
   - What is your class standing in college?
   - Where are you from?
2. Why did you choose to study/perform research in astronomy?
   - What were early influences in life to interest you in astronomy/science?
   - Do your parents have who work in a STEM field?
   - Did your parents encourage you to pursue STEM?
   - With whom do you perform research?
   - Did your research mentor influence your decision to study astronomy?
   - What high school did you go to? Did you go to a community college?
   - When did you declare astronomy major or decide to join a research group in astronomy?
   - What area do you do research in?
   - Do you enjoy it? Why?
   - What topic(s) do you like most in astronomy? Why do you like that topic?/ What do you find interesting about the topic?

- How do you think your gender/ethnicity experiences or intersectional identities may have influenced your decision to study astronomy?
- Do you feel confident in your ability to do work related to astronomy?
- Does doing research change your opinion about astronomy/physics?
- Describe your interactions with your peers and mentors.
- How do your interactions with peers and mentors affect your sense of identity as an astronomer?
- How have your past experiences in science influenced your interest in astronomy.
- Do you see yourself as an astronomer or "astronomy person"?
- Do feel a sense of community/belonging in the astronomy community/department? How does this affect your identification with being an astronomer?

3. Do you think the physics and/or astronomy labs helped with your understanding of the scientific method?
    - Did the physics/astronomy labs improve the way you think critically and improve your scientific literacy?
    - Did the physics/astronomy labs help with your understanding of astronomy concepts?
    - Did the physics/astronomy labs help with your understanding of physics concepts?
    - After your introductory physics lab, did your opinion changed about astronomy?
4. Do you enjoy astronomy more or less than before you enrolled in the introductory physics lab?
    - Why? What may have changed your mind?
5. To what extent did you interact with your instructors or other students in your physics/astronomy labs?
    - Did you go to office hours?
    - To what extent did you socialize with your peers?
    - Whom did you contact if you didn't understand how to complete the lab?
    - Do you think you would've benefited from formal discussion groups? (if taking online labs).
6. What are your plans after graduating?